\documentclass[12pt,a4paper]{article}
\usepackage{pdflscape}
\usepackage[top=1in, bottom=1in, left=1in, right=1in]{geometry}
\usepackage{rotating}
\usepackage{enumitem}
\setlist[description]{font=\ttfamily,leftmargin=2cm,style=nextline}
\usepackage{amsmath,amssymb}
\usepackage{graphicx}
\usepackage{color}
\usepackage{dcolumn}
\usepackage{bm}
\usepackage[numbers,super,comma,sort&compress]{natbib}

\usepackage[version=3]{mhchem} 
\usepackage{hyperref}

\usepackage[binary-units=true]{siunitx}

\usepackage{xcolor,listings}
\usepackage{caption}
\usepackage{subcaption}
\author{Taweetham Limpanuparb
\thanks{Mahidol University International College, Mahidol University, Nakhonpathom 73170, Thailand 
Corresponding author: taweetham.lim@mahidol.ac.th}, %
Josh Milthorpe\thanks{IBM T.J. Watson Research Center, P.O. Box 704, Yorktown Heights, New York 10598, USA
} 
}

\title{Associated Legendre Polynomials and \\Spherical Harmonics Computation \\for Chemistry Applications} 

\begin{document}

\maketitle

\begin{abstract}
Associated Legendre polynomials and spherical harmonics are central to calculations in many fields of science and mathematics -- not only chemistry but computer graphics, magnetic, seismology and geodesy. 
There are a number of algorithms for these functions published since 1960 but none of them satisfy our requirements.
In this paper, we present a comprehensive review of algorithms in the literature and, based on them, propose an efficient and accurate code for quantum chemistry. 
Our requirements are to efficiently calculate these functions for all non-negative integer degrees and orders up to a given number ($\leq 1000$) and the absolute or the relative error of each calculated value should not exceed $10^{-10}$.
We achieve this by normalizing the polynomials, employing efficient and stable recurrence relations, and precomputing coefficients.
The algorithm presented here is straightforward and may be used in other areas of science.
\end{abstract}

\section{Introduction}

In 1782, Legendre\cite{legendre1785} introduced polynomials $P_\ell$ as the coefficients in the expansion of the Newtonian potential
\begin{equation}
\frac{1}{r_{12}} = \frac{1}{|\bm{r}_1-\bm{r}_2|}=\sum_{l=0}^\infty \frac{r_<^l}{r_>^{l+1}}P_\ell(\cos \theta)
\end{equation}
where $r_<=\min\left(|\bm{r}_1|,|\bm{r}_2|\right)$, $r_>=\max\left(|\bm{r}_1|,|\bm{r}_2|\right)$ and $\bm{r}_1 \cdot \bm{r}_2 = |\bm{r}_1| |\bm{r}_2|\cos \theta$.  
Associated Legendre Polynomials (ALPs)\footnote{ALPs are sometimes referred to as Associated Legendre Functions (ALFs)\cite{} because the $(1-x^2)^{m/2}$  factor is not a polynomial for odd $m$. 
This does not necessarily mean Associated Legendre functions of the second kind, $Q_\lambda^\mu$.} 
$P_\ell^m$ of degree $\ell$ and order $m\geq0$ may be defined as the $m^\text{th}$ derivative of $P_\ell$,
\begin{equation}
P_\ell^{m}(x) = (-1)^m\ (1-x^2)^{m/2}\ \frac{d^m}{dx^m}P_\ell(x).  \label{eq:ALP1}
\end{equation} The negative order can be related to the corresponding positive order via a proportionality constant that involves only $\ell$ and $m$,
\begin{equation}
P^{-m}_\ell(x) = (-1)^m \frac{(\ell-m)!}{(\ell+m)!} P^{m}_\ell(x). \label{eq:ALP2}
\end{equation}
These ALPs are closely related to the spherical harmonics (SHs)\footnote{There are several alternative definitions of ALPs and SHs which are commonly used in the fields of magnetic, geodesy and seismology.  
They are slightly different e.g. presence or absence of $(-1)^m$ or other normalization factor.  
Our algorithm may be easily modified to suit their definitions.}
\begin{equation}
Y_\ell^m(\theta, \phi) =  \sqrt{\frac{(2\ell+1)(\ell-m)!}{4\pi(\ell+m)!}}\ P_\ell^{m}(\cos \theta)\ e^{im\phi}
\end{equation}
which are the analytic solutions for wavefunctions of the hydrogen atom and are common ingredients for some quantum chemistry calculations.\cite{RO1, RO2, RO3, RO4, RO5, RO6, RO7, RO8, ROThesis}  
It follows from the above equations that $P_\ell^{m}$ are the essential part for the computation of $Y_\ell^m$.  
In this manuscript, we propose an algorithm to accurately and efficiently compute ALPs and SHs in the context of quantum chemical applications.
We note that the argument $x = \cos \theta$ is real and bounded $|x| \leq 1$ and both degree $\ell$ and order $m$ are integers satisfying the conditions $-\ell \leq m \leq \ell$ and $\ell\geq0$.

\section{Review of algorithms for ALPs and SHs}
The standard approach for special function calculation in quantum chemistry software is to use recurrence relations (RRs).  
However, there are a number of aspects that need to be taken into consideration.  
There are myriad of RRs but not all of them are practical for computation using floating-point arithmetic.  
For some RRs, round-off error may rapidly propagate and become significant if used in a certain direction. 
When the same RR is applied in the reverse direction the numerical behavior might be the opposite.
This is the basis of Miller's backward algorithm. \cite{gautschi1967, olver1972, olver1964, lordRayleigh1910}

Since the ultimate purpose of calculation is to obtain SHs, the ALPs may also be modified or normalized to improve the stability of the RRs.  
ALPs and SHs have been a subject of numerous publications since 1960, but the best approach for ALP calculation may differ depending on the application.   
``Numerical Recipes'', one of the most famous books on algorithms and numerical analysis, has even presented different algorithms in its second and third editions.\cite{NR2nd1992, NR3rd2007}  
The list below gives a brief summary of the algorithms developed over the past five decades ordered by year of publication. 
We indicate whether $P_\ell^m$ are normalized, however, normalization schemes differ between publications.

\begin{description}
\item[\textbf{Ref}]{\textbf{Description}}
\item[\citenum{galler1960}] {Calculation of $P_\ell$ by a RR in the direction of increasing $\ell$}
\item[\citenum{herndon1961}] {Calculation of $P_\ell^m$ for $\ell<20$ \\ 
Only half-page source code was provided in the manuscript.  }
\item[\citenum{wiggins1971}] {Calculation of $P_\ell^m$ tested up to $\ell=m=50$ \\
Start from $P_\ell^\ell$ and use a RR in the direction of decreasing $m$.}
\item[\tiny\citenum{braithwaite1973, NR2nd1992, schneider2010}]{Calculation of $P_\ell^m$\\
Start from $P_\ell^\ell$ and use a RR in the direction of increasing $\ell$.}
\item[\citenum{smith1981, lozier1981}]{Calculation of normalized $P_\ell^m$ using extended-range arithmetic (up to $\ell=m=10000$)\\
Start from normalized $P_\ell^\ell$ and use a RR in the direction of decreasing $m$.}
\item[\citenum{olver1983}]{Calculation of $P_\ell^m$ using extended-range arithmetic (up to $\ell=m=100000$) \\
Start from $P_\nu^m$ and $P_{\nu+1}^m$ obtained by series expansion and use a RR in the direction of increasing $\ell$. 
If $P_\ell^m$ for a range of $m$ are needed, use a RR in the direction of increasing $m$ when $m\leq0$ and use reflection formula to relate $P_\ell^m$ and $P_\ell^{-m}$.}
\item[\citenum{libbrecht1985}]{Calculation of normalized $P_\ell^m$ \\
Start from normalized $P_\ell^\ell$ and use a RR in the direction of decreasing $m$. 
Alternatively, use an L-shape RR involving $P_\ell^{m-1}$, $P_\ell^m$ and $P_{\ell+1}^m$ in the direction of increasing $\ell$ for larger degree and order.}
\item[\citenum{holmes2002}]{Calculation of normalized $P_\ell^m$, review of existing methods for Clenshaw summation used in geodesy}
\item[\citenum{jekeli2007}]{Calculation of normalized $P_\ell^m$ using extended-range arithmetic\\
Start from $\ell=0$, use a RR in the direction of increasing $\ell$ and finally use another RR in the direction of increasing $m$ with a cut-off value of $m$ to set the result to 0. }
\item[\citenum{NR3rd2007}]{Calculation of normalized $P_\ell^m$\\
Start from normalized $P_\ell^\ell$ and use a RR in the direction of increasing $\ell$. 
The approach is similar to Refs \citenum{braithwaite1973, NR2nd1992, schneider2010} for the direction of the RR and similar to Refs \citenum{smith1981, lozier1981} for normalization.}
\item[\citenum{GNU2009}]{Calculation of normalized $P_\ell^m$\\
Start from normalized $P_\ell^\ell$ and uses a RR in the direction of increasing $\ell$.  
Compute all $P_\ell^m$ in the $\ell$ direction for a fixed $m$ value.
The approach is similar to Ref\citenum{NR3rd2007}.}
\item[\citenum{BOOST2011}]{Calculation of $P_\ell^m$\\
Use RR in the direction of increasing $\ell$ and return a single value of $P_\ell^m$ only. 
Extensive accuracy test for $\ell\leq120$ is published on boost.org website and the author claims that other libraries produce identical error rates.
The function is unlikely to produce sensible results for $\ell>120$. }  
\item[\citenum{sloan2013}]{Calculation of real $Y_{\ell,m}$\\
GPU-optimized code using precomputed, hard-coded coefficients.
Benchmark results presented for order $\leq 10$.}
\end{description}

\section{Description of our Algorithm}

\subsection{Motivation and working equation}

Our target applications are resolutions of the Coulomb operator in quantum chemistry codes~\cite{RO1, RO2, RO3, RO4, RO5, RO6, RO7, RO8, ROThesis}.
For these resolutions, moderate values of degree and order ($\ell \leq 1000$) are required for $Y_{\ell,m}$ and it is imperative that the program should run efficiently on standard double-precision architecture.
The target accuracy is a relative error of less than $10^{-10}$.
However, for extremely small values of $Y_{\ell,m}$, relative error may be high but irrelevant to the final chemistry result.  
Therefore, if a calculated $Y_{\ell,m}$ fails the relative error test, it may still be acceptable if the absolute error is less than $10^{-10}$.  
This is because the value of $Y_{\ell,0}$ are bound by $\sqrt{(2\ell+1)/4\pi}$ which grows slowly with respect to $\ell$.

For many applications, real spherical harmonics
\begin{equation}
Y_{\ell,m}(\theta,\phi)=\begin{cases}
 \bar{P}_\ell^{|m|} (\cos \theta) \sin |m|\phi & \text{if}\ m<0\\
 \bar{P}_\ell^0 (\cos \theta)/\sqrt{2} & \text{if}\ m=0\\
 \bar{P}_\ell^m (\cos \theta) \cos m\phi  & \text{if}\ m>0.
\end{cases}
\end{equation}
are preferred to complex $Y_{\ell}^m$ as the real ones require less storage space and are computationally cheaper to generate and use.  
It follows from the above equation that our choice of normalization is
\begin{equation}
\bar{P}_\ell^m=\sqrt{\frac{(2l+1) (l-m)!}{2\pi(l+m)!}}P_\ell^m
\end{equation}
where ALPs of only $m\geq0$ are required to generate whole set of real SHs.

We have used the algorithms in the previous section and found that the approach described in the third edition of ``Numerical Recipes''\cite{NR3rd2007} and GNU Scientific Library\cite{GNU2009} is the most appropriate for our application.  
Normalization of $P_\ell^m$ not only helps the conversion to $Y_{\ell,m}$ but also results in better numerical stability. 
Our ALP and SH algorithms are based on the following design principles:
\begin{itemize}
\item Normalize $P_\ell^m$ to avoid overflow/underflow.
\item Use a RR in the direction of increasing $\ell$ for ALPs for stability.
\item Use trigonometric RRs for $\sin$ and $\cos$ functions in SHs to save time.
\item Precompute coefficients in the RRs to reduce computational cost.
\item Compute an entire set of normalized $P_\ell^m$ where $m\geq0$ in a single function call to save overhead cost.
\item Avoid loop dependencies in inner loops, allowing operations to be vectorized and pipelined for execution.
\end{itemize}
The set of working equations for our algorithm is described below.
\begin{align}
a_\ell^m&=\sqrt{\frac{4l^2-1}{l^2-m^2}}\label{eq:coeff_a} \\
b_\ell^m&=-\sqrt{\frac{(l-1)^2-m^2}{4(l-1)^2-1}}\label{eq:coeff_b}\\
\bar{P}_0^0 &= \sqrt{\frac{1}{2\pi}} \\
x&=\cos \theta \\
y&=\sin \theta \\
\bar{P}_m^m &= -\sqrt{1+\frac{1}{2m}}\ y \ \bar{P}_{m-1}^{m-1}\\
\bar{P}_{m+1}^m &= \sqrt{2m+3}\ x \bar{P}_m^m \\
\bar{P}_\ell^m &= a_\ell^m (x\bar{P}_{\ell-1}^m + b_\ell^m \bar{P}_{\ell-2}^m)
\end{align}
To reduce the computational cost, two-term RRs
\begin{align}
\cos m\phi &= 2 \cos \phi \cos (m-1)\phi - \cos (m-2)\phi \label{eq:one}\\
\sin m\phi &= 2 \cos \phi \sin (m-1)\phi - \sin (m-2)\phi
\end{align}
or one-term RRs
\begin{align}
\cos (m\phi) &= \cos \psi - [\alpha \cos \psi + \beta \sin \psi]  \\
\sin (m\phi) &= \sin \psi - [\alpha \sin \psi - \beta \cos \psi] \\
\alpha &= 2 \sin^2 \left(\frac{\phi}{2} \right)\\
\beta &= \sin \phi \\
\psi &= (m-1)\phi \label{eq:two}
\end{align}
may also be used to calculate the sinusoidal functions in SHs. 
Using the RRs, $\sin \phi$ and $\cos \phi$ are the only two expensive transcendental function operations required to generate the whole set of SHs.

\subsection{Implementation}

\subsubsection*{Initialization}
The coefficients $a_\ell^m$~\eqref{eq:coeff_a} and $b_\ell^m$~\eqref{eq:coeff_b} are precomputed for all $\ell \leq L$,$m \leq \ell$ as follows:
\begin{lstlisting}[language=c]
#define PT(l,m) ((m)+((l)*((l)+1))/2)

for (size_t l=2; l<=LL; l++) {
  double ls=l*l, lm1s = (l-1)*(l-1); 
  for (size_t m=0; m<l-1; m++) {
    double ms=m*m;
    A[PT(l,m)] = sqrt((4*ls-1.)/(ls-ms));
    B[PT(l,m)] = -sqrt((lm1s-ms)/(4*lm1s-1.));
  }
}
\end{lstlisting}

\subsubsection*{$P_\ell^m$}
The function \lstinline!computeP! computes an entire set of $\bar{P}_\ell^m(x)$ and stores in the array \lstinline!P!.
\begin{lstlisting}[language=c]
void computeP(const size_t L, 
              const double* const A, const double* const B, 
              double* const P, const double x) {
  const double sintheta = sqrt(1.-x*x);
  double temp = 0.39894228040143267794; // = sqrt(0.5/M_PI)
  P[PT(0,0)] = temp;
  if (L > 0) {
    const double SQRT3 = 1.7320508075688772935;
    P[PT(1,0)] = x*SQRT3*temp;
    const double SQRT3DIV2 = -1.2247448713915890491;
    temp = SQRT3DIV2*sintheta*temp;
    P[PT(1,1)] = temp;

    for (size_t l=2; l<=L; l++) {
      for (size_t m=0; m<l-1; m++) {
        P[PT(l,m)] = A[PT(l,m)]*(x*P[PT(l-1,m)]
                     + B[PT(l,m)]*P[PT(l-2,m)]);
      }
      P[PT(l,l-1)] = x*sqrt(2*(l-1)+3)*temp;
      temp = -sqrt(1.0+0.5/l)*sintheta*temp;
      P[PT(l,l)] = temp;
    }
  }
}
\end{lstlisting}

\subsubsection*{$Y_{l,m}$}
The function \lstinline!computeY! computes an entire set of $Y_{\ell,m}(\theta,\phi)$ and stores in the array \lstinline!Y!.
\begin{lstlisting}[language=c]
#define YR(l,m) ((m)+(l)+((l)*(l)))

void computeY(const size_t L, const double * const P, 
              double * const Y, const double phi) {
  for (size_t l=0; l<=L; l++)
    Y[YR(l,0)] = P[PT(l,0)] * 0.5 * M_SQRT2;

  double c1 = 1.0, c2 = cos(phi);
  double s1 = 0.0, s2 = -sin(phi);
  double tc = 2.0 * c2;
  for (size_t m=1; m<=L; m++) {
    double s = tc * s1-s2;
    double c = tc * c1-c2;
    s2 = s1; s1 = s; c2 = c1; c1 = c;
    for (size_t l=m; l<=L; l++) {
      Y[YR(l,-m)] = P[PT(l,m)] * s;
      Y[YR(l,m)]  = P[PT(l,m)] * c;
    }
  }
}
\end{lstlisting}

\section{Numerical results}

\subsection{Accuracy}
We used Mathematica to calculate reference values of ALPs.
The ALPs are generated symbolically first and are evaluated at the final stage using extended precision arithmetic (100 digits for $\ell \leq 
100$ and 1000 digits for $\ell=1000$).
We then compare the calculated and reference values by measuring the magnitudes of the value, the absolute error, and the relative error:
\begin{align}
\chi &= \max \left(\log_{10}\left| X \right|, -324 \right) \\
\epsilon_\text{a} &= \max \left(\log_{10}\left| X_\text{cal}-X_\text{ref} \right|, -324 \right) \\
\epsilon_\text{r} &= \min \left( \max \left( \log_{10}\left| \frac{X_\text{cal}}{X_\text{ref}}-1  \right|, -16 \right), 0 \right)
\end{align}
where $X$ stands for the value of $\bar{P}_\ell^m$, $\sin m\phi$, or $\cos m\phi$.
To avoid problematic values, the magnitude of the relative error $\epsilon_\text{r}$ is calculated only when $X_\text{ref}\neq0$ and
max and min are used in the definitions.
 
\begin{landscape}
\begin{figure}
\begin{center}
\includegraphics[width=250mm,trim=0mm 10mm 5mm 10mm,clip=true]{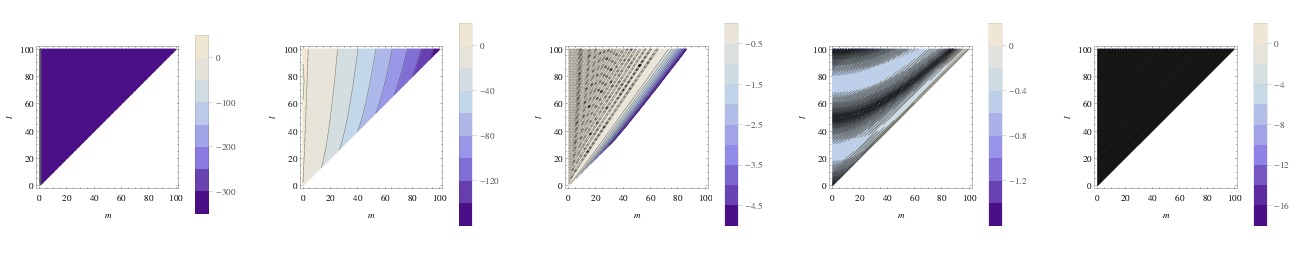}\\
a. magnitude of calculated values, $\chi_\text{cal}$ and reference values, $\chi_\text{ref}$ (Both are visually identical.)\\
\includegraphics[width=250mm,trim=0mm 10mm 5mm 10mm,clip=true]{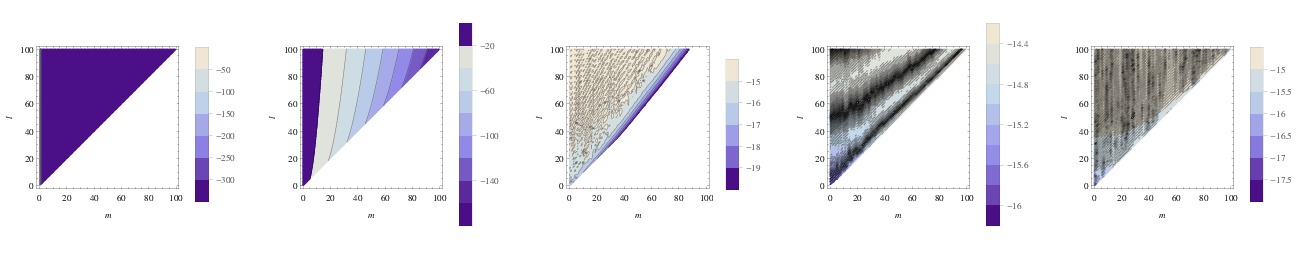}\\
b. magnitude of absolute errors, $\epsilon_\text{a}$\\
\includegraphics[width=250mm,trim=0mm 10mm 5mm 10mm,clip=true]{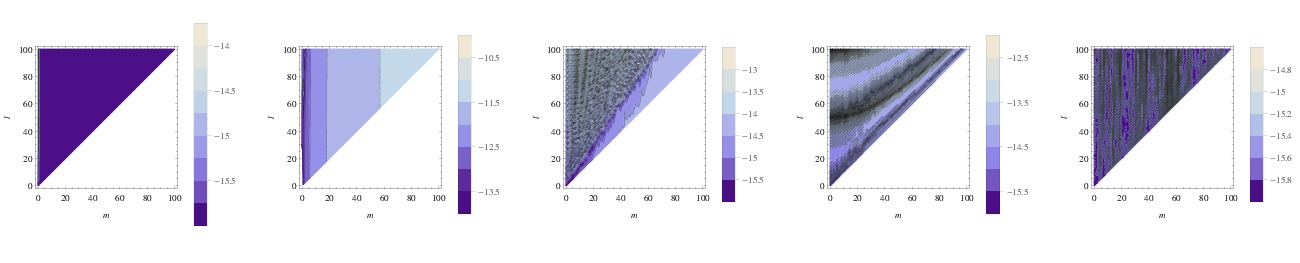}\\
c. magnitude of relative errors, $\epsilon_\text{r}$
\caption{Accuracy investigation of $\bar{P}_\ell^m$ for $l \le 100$ for $\theta=0, \frac{\pi}{100}, \frac{\pi}{4}, \frac{49\pi}{100}, \frac{\pi}{2}$ from left to right respectively\label{fig:plm100}}
\end{center}
\end{figure}
\end{landscape}

We first investigate the accuracy of $\bar{P}_\ell^m$ where $\ell\leq100$.  
Figure~\ref{fig:plm100} shows a representative example for $\theta=0, \frac{\pi}{100}, \frac{\pi}{4}, \frac{49\pi}{100}, \frac{\pi}{2}$.  
The graphical representations of reference and calculated values perfectly match and confirm earlier findings\cite{jekeli2007} that $\bar{P}_\ell^m$ diminish rapidly with respect to $m$ when $\left| \cos \theta \right|$ is close to 1 and oscillatory when $\left| \cos \theta \right|$ is close to 0.
The magnitude of absolute and relative errors are well below the target accuracy of $10^{-10}$.

\begin{figure} 
    \begin{subfigure}{0.5\textwidth}
    \includegraphics[width=78mm]{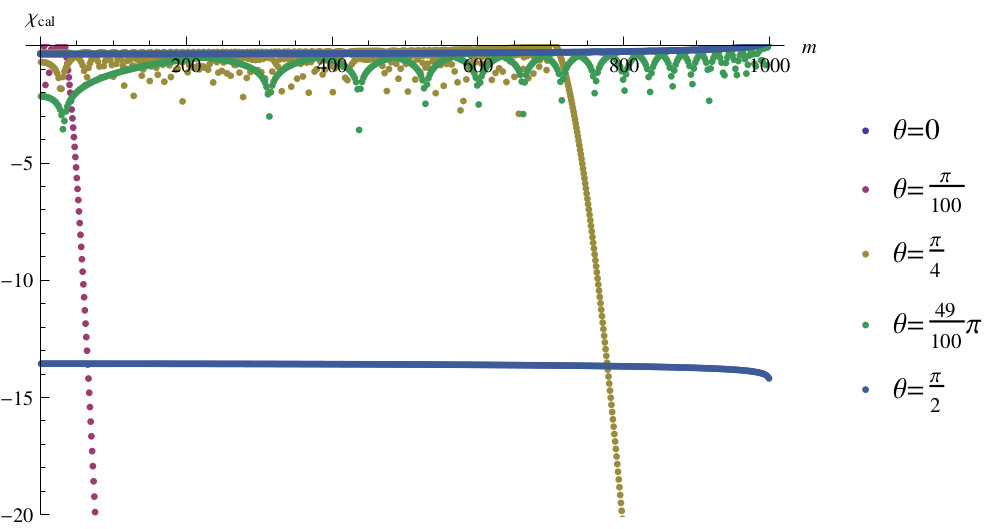}
    \end{subfigure}
    \begin{subfigure}{0.5\textwidth}
    \includegraphics[width=78mm]{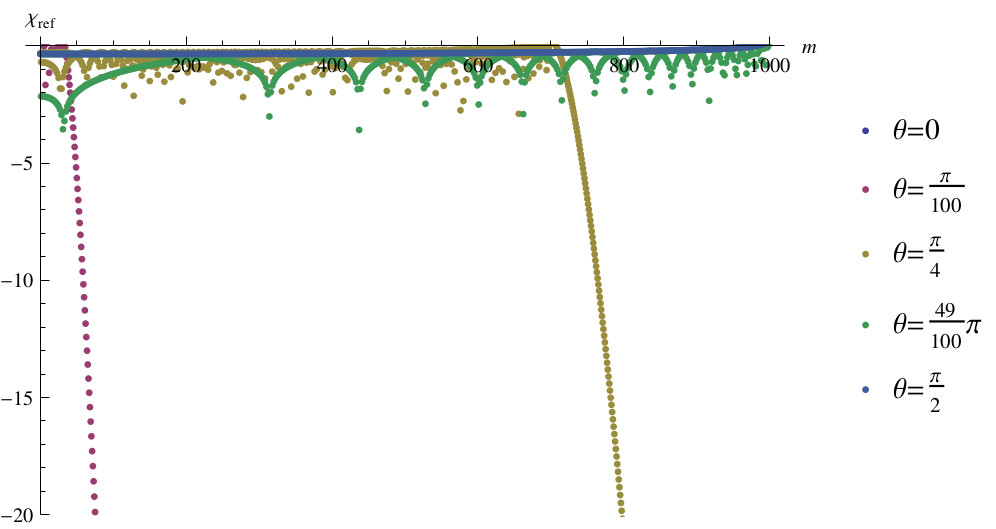}
    \end{subfigure}

    \begin{subfigure}{0.5\textwidth}
    \includegraphics[width=78mm]{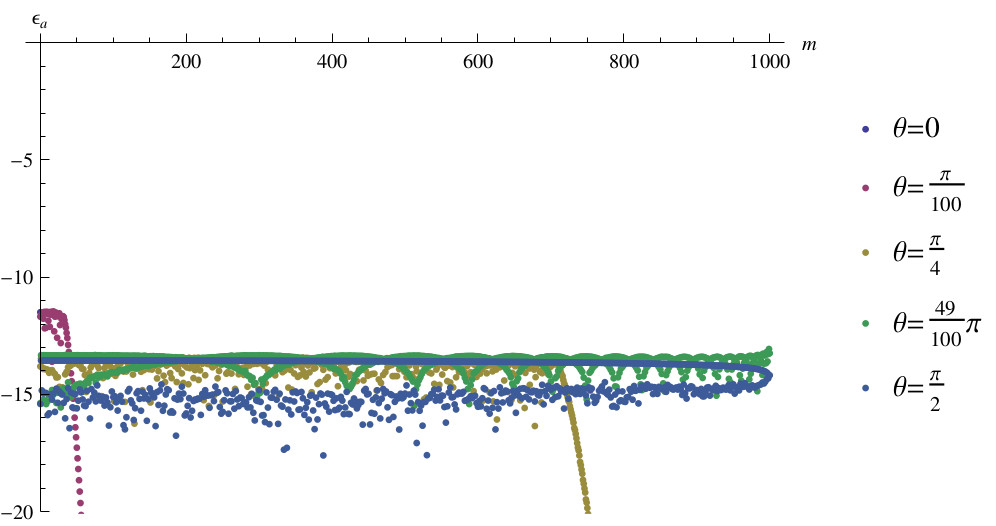}
    \end{subfigure}
    \begin{subfigure}{0.5\textwidth}
    \includegraphics[width=78mm]{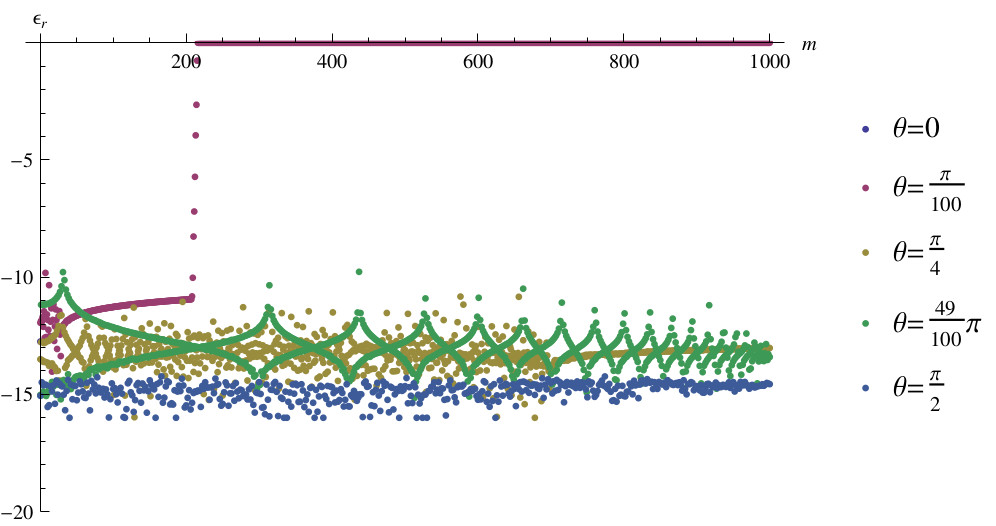}
    \end{subfigure}
\caption{Magnitude of calculated and reference values of $\bar{P}_{1000}^m$ and their errors\label{fig:plm1000}}
\end{figure}

We now consider $\ell=1000$ case in Figure~\ref{fig:plm1000}.  
The magnitude of $\bar{P}_\ell^m$ decreases rapidly for small $\theta$.  
For $\theta=\pi/4$ the value of $\bar{P}_\ell^m$ diminishes with higher $m$ and for $\theta=49\pi/50,\pi/2$ we observe that $\bar{P}_\ell^m$ are oscillatory and remains significant until $m=1000$. 
We note that for $\theta=\pi/2$, $\bar{P}_\ell^m=0$ when $m$ is odd and the numerical noise in this case can be seen clearly in the graph of calculated $\bar{P}_\ell^m$.  
However, the noise is small in magnitude and the absolute and relative error are well below $10^{-10}$. 

Finally, we confirm that the RRs for sinusoidal functions are sufficiently accurate for our purpose.
We wrote a stand-alone C program that employs the one-term and two-terms RRs mentioned in \eqref{eq:one} to \eqref{eq:two} in the direction of increasing $m$. 
The program was run for $0 \leq m\leq 1000$ and $\phi=0.000000, 0.000001, 0.000002, ..., 6.283185$.
Table~\ref{tab:sincosRR} shows the maximum error for the two sets of RRs. 
Though the relative error may appear to be high, our additional investigation ascertains that those cases occur when the absolute values are low.
A further investigation shows that we do not find any cases where both absolute and relative errors are above $10^{-10}$ for two-term RRs and $10^{-12}$ for one-term RRs.

\begin{table}
\begin{tabular}{| l|r r |r r| r r| r r|}
\hline
RRs & \multicolumn{2}{c|}{$\epsilon_\text{a}(\sin)$} & \multicolumn{2}{c|}{$\epsilon_\text{a}(\cos)$} & \multicolumn{2}{c|}{$\epsilon_\text{r}(\sin)$} &  \multicolumn{2}{c|}{$\epsilon_\text{r}(\cos)$} \\
  & max & average & max & average & max & average & max & average \\
\hline
one-term   &-12.13 & -13.29  & -12.13 & -13.29 & 0.00 & -13.41 & 0.00 & -13.41  \\
two-term   &-10.59 & -13.16 & -10.20  & -13.12 & 0.00 & -13.43& 0.00 & -13.43 \\
\hline
\end{tabular}
\caption{Absolute and relative errors from one- and two-term RRs for sine and cosine functions. For absolute error, we average the term $\left| X_\text{cal}-X_\text{ref} \right|$ first and then take log but for relative error we find $\epsilon_\text{r}$ first and then find the average.}
\label{tab:sincosRR}
\end{table}

The error analysis for SHs is straightforward.
For example, absolute error of SHs for $m>0$ can be expanded as
\begin{align}
\left(\bar{P}^{m}_\ell
+\Delta_P\right)\left(\cos m\phi + \Delta_c\right) &= Y_{\ell,m}
+\Delta_P \cos m\phi +\Delta_c \bar{P}^{m}_\ell
\\
\Delta_Y &= \Delta_P \cos m\phi +\Delta_c \bar{P}^{m}_\ell
+\Delta_P\Delta_c.
\end{align}
It is obvious that the last term $\Delta_P\Delta_c$ is negligible and the behavior of the other two error terms are more or less predictable since $\Delta$ is multiplied to a bounded function.
The largest $\bar{P}^{m}_\ell$ in our case is $\bar{P}^{0}_{1000}(1)=\sqrt{\frac{2001}{2\pi}}\approx18$.
Since absolute errors, $\Delta_P$ and $\Delta_c$ are well below $10^{-10}$ we conclude that the resulting $\Delta_Y$ is also below this threshold provided that one-term RRs are used for the trigonometry functions.
A similar analysis is also applicable to $m=0$ (change $\cos m\phi$, $\Delta_c$ to $1/\sqrt{2}$, 0) and $m<0$ (change $\cos m\phi$, $\Delta_c$ to $\sin m\phi$, $\Delta_s$).

From the analysis here, it is anticipated that our algorithm may be used beyond $\ell=1000$.
However, trigonometric RRs may be less attractive as the computational cost saving is no longer significant.  
A cut-off scheme for small values may be more helpful in this circumstance.
We do not provide an analysis for this as the reference values are difficult to calculate and it is beyond the scope of our chemical applications.

\subsection{Computational cost}

\begin{figure}[hbtp]
    \begin{subfigure}{0.5\textwidth}
        \hspace{-1cm}
        \includegraphics[width=9cm,trim=0mm 10mm 0mm 10mm, clip=true]{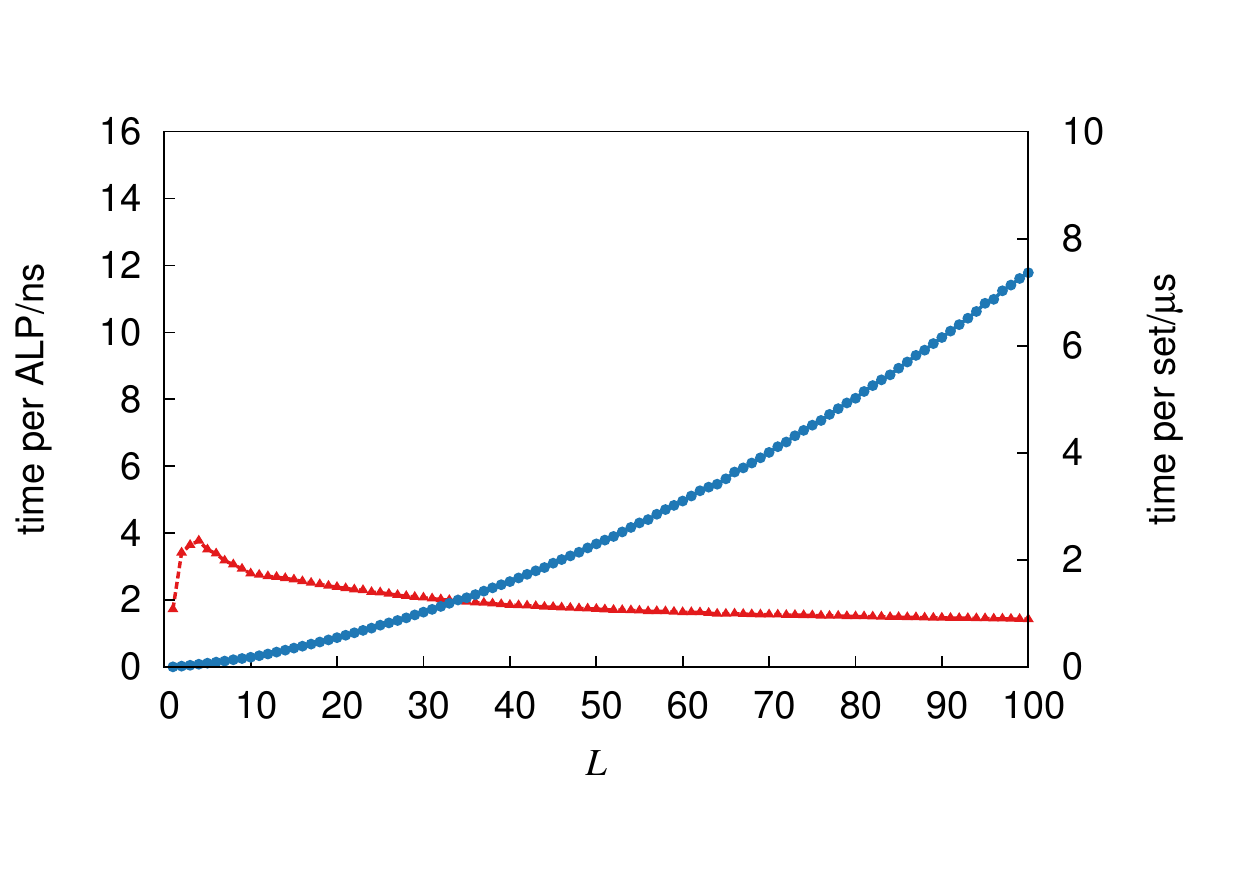}
        \caption{Core-i7}
    \end{subfigure}
    \begin{subfigure}{0.5\textwidth}
        \includegraphics[width=9cm,trim=0mm 10mm 0mm 10mm, clip=true]{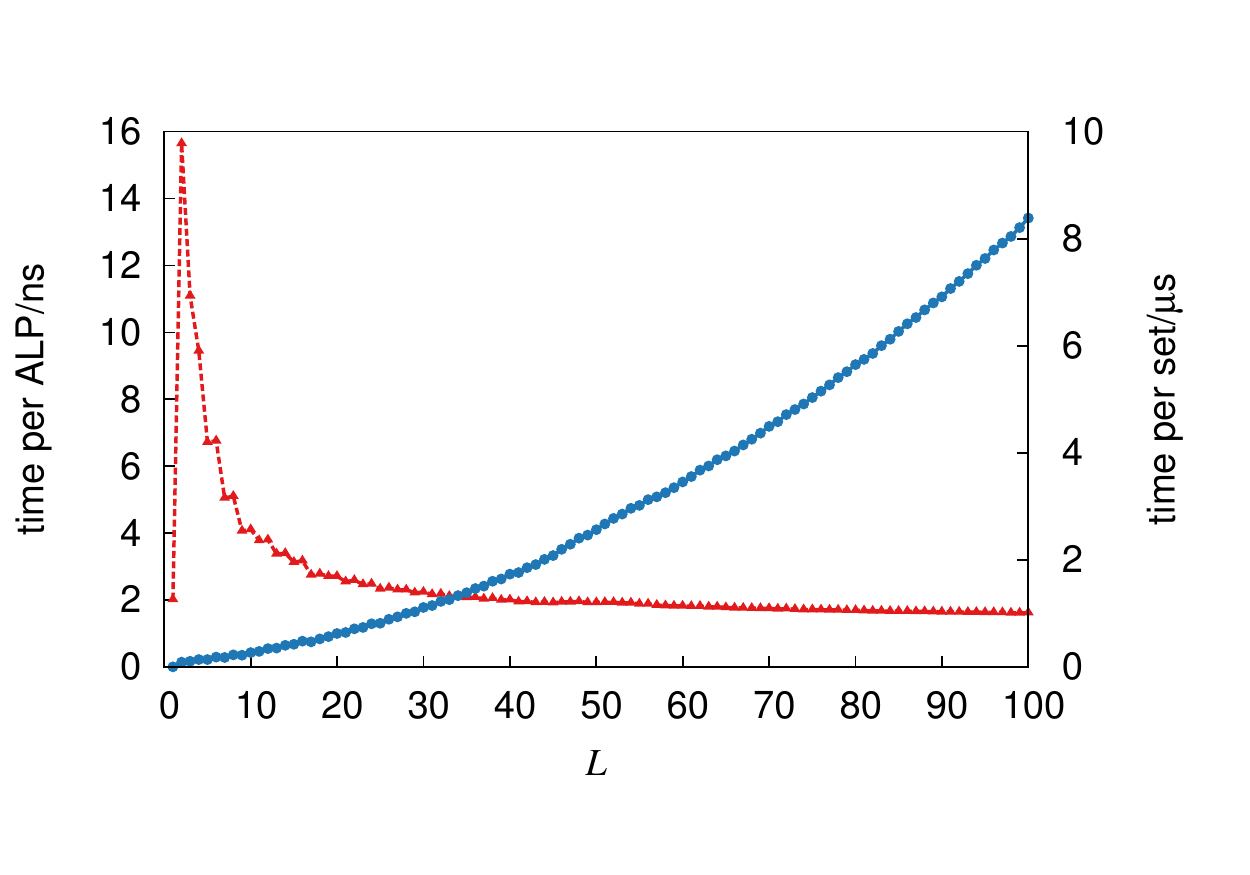}
        \caption{POWER7}
    \end{subfigure}
    \caption{Time to compute $\bar{P}_\ell^m$ for maximum degree $L \le 100$.}
    \label{fig:plm_cost}
\end{figure}

We measured the performance of our \lstinline!computeP! function, which computes all $\bar{P}_\ell^m$ where $0 \leq m \leq \ell \leq L$.
Single-core timings were measured on two different computing platforms: an Intel Ivy Bridge Core-i7-3740QM @ \SI{2.7}{\giga \Hz}, and an IBM POWER7 @ \SI{3.8}{\giga \Hz}.
Figure~\ref{fig:plm_cost} shows the average time to compute one $\bar{P}_\ell^m$, as well as the time to compute an entire set of $\bar{P}_\ell^m$ for $\theta = \pi / 20$ 
for maximum degree $L \le 100$.

On both platforms, time per $\bar{P}_\ell^m$ is highest for small values of $L$.
On the Core-i7\footnote{On Core-i7, the GNU C++ compiler was used with optimization level \lstinline!-O3!.},
time per $\bar{P}_\ell^m$ decreases from \SI{3.8}{\nano \second} for $(L=4)$ to \SI{1.4}{\nano \second} for $(L=100)$, which is approximately 3.8 cycles at a clock speed of \SI{2.7}{\giga \Hz}.
On the POWER7\footnote{On POWER7, the IBM xlC compiler was used with optimization level \lstinline!-O5!.},
time per $\bar{P}_\ell^m$ decreases from \SI{15.6}{\nano \second} for $(L=2)$ to \SI{1.6}{\nano \second} for $(L=100)$, which is approximately 6.2 cycles at a clock speed of \SI{3.8}{\giga \Hz}.
Larger values of $L$ benefit from long inner loops which can be effectively vectorized and pipelined for execution.
The time required for initialization is on the order of \SI{100}{\micro \second} for $L=100$, which is insignificant when amortized over a large number of calls to \lstinline!computeP!. 

\section{Concluding remarks}
We have proposed an algorithm for the calculation of ALPs and SHs.  
Accuracy analysis was conducted for degree and order up to 1000 and found that absolute or relative error are satisfactorily below $10^{-10}$.
Timing experiments showed that our C++ implementation takes less than four cycles on average to produce an ALP.
This new code will be used in our future quantum chemistry work.




\small
\bibliography{ALP.bib}

\providecommand*{\mcitethebibliography}{\thebibliography}
\csname @ifundefined\endcsname{endmcitethebibliography}
{\let\endmcitethebibliography\endthebibliography}{}
\begin{mcitethebibliography}{30}
\providecommand*{\natexlab}[1]{#1}
\providecommand*{\mciteSetBstSublistMode}[1]{}
\providecommand*{\mciteSetBstMaxWidthForm}[2]{}
\providecommand*{\mciteBstWouldAddEndPuncttrue}
  {\def\EndOfBibitem{\unskip.}}
\providecommand*{\mciteBstWouldAddEndPunctfalse}
  {\let\EndOfBibitem\relax}
\providecommand*{\mciteSetBstMidEndSepPunct}[3]{}
\providecommand*{\mciteSetBstSublistLabelBeginEnd}[3]{}
\providecommand*{\EndOfBibitem}{}
\mciteSetBstSublistMode{f}
\mciteSetBstMaxWidthForm{subitem}
{(\emph{\alph{mcitesubitemcount}})}
\mciteSetBstSublistLabelBeginEnd{\mcitemaxwidthsubitemform\space}
{\relax}{\relax}

\bibitem[Legendre(1785)]{legendre1785}
A.-M. Legendre, \emph{M{\'e}moires de Math{\'e}matiques et de Physique,
  pr{\'e}sent{\'e}s {\`a} l'Acad{\'e}mie royale des sciences (Paris)}, 1785,
  \textbf{10}, 411--435\relax
\mciteBstWouldAddEndPuncttrue
\mciteSetBstMidEndSepPunct{\mcitedefaultmidpunct}
{\mcitedefaultendpunct}{\mcitedefaultseppunct}\relax
\EndOfBibitem
\bibitem[Varganov \emph{et~al.}(2008)Varganov, Gilbert, Deplazes, and
  Gill]{RO1}
S.~A. Varganov, A.~T.~B. Gilbert, E.~Deplazes and P.~M.~W. Gill, \emph{J. Chem.
  Phys.}, 2008, \textbf{128}, 201104\relax
\mciteBstWouldAddEndPuncttrue
\mciteSetBstMidEndSepPunct{\mcitedefaultmidpunct}
{\mcitedefaultendpunct}{\mcitedefaultseppunct}\relax
\EndOfBibitem
\bibitem[Gill and Gilbert(2009)]{RO2}
P.~M.~W. Gill and A.~T.~B. Gilbert, \emph{Chem. Phys.}, 2009, \textbf{356},
  86--90\relax
\mciteBstWouldAddEndPuncttrue
\mciteSetBstMidEndSepPunct{\mcitedefaultmidpunct}
{\mcitedefaultendpunct}{\mcitedefaultseppunct}\relax
\EndOfBibitem
\bibitem[Limpanuparb and Gill(2009)]{RO3}
T.~Limpanuparb and P.~M.~W. Gill, \emph{Phys. Chem. Chem. Phys.}, 2009,
  \textbf{11}, 9176--9181\relax
\mciteBstWouldAddEndPuncttrue
\mciteSetBstMidEndSepPunct{\mcitedefaultmidpunct}
{\mcitedefaultendpunct}{\mcitedefaultseppunct}\relax
\EndOfBibitem
\bibitem[Limpanuparb \emph{et~al.}(2011)Limpanuparb, Gilbert, and Gill]{RO4}
T.~Limpanuparb, A.~T.~B. Gilbert and P.~M.~W. Gill, \emph{J. Chem. Theory
  Comput.}, 2011, \textbf{7}, 830--833\relax
\mciteBstWouldAddEndPuncttrue
\mciteSetBstMidEndSepPunct{\mcitedefaultmidpunct}
{\mcitedefaultendpunct}{\mcitedefaultseppunct}\relax
\EndOfBibitem
\bibitem[Limpanuparb and Gill(2011)]{RO5}
T.~Limpanuparb and P.~M.~W. Gill, \emph{J. Chem. Theory Comput.}, 2011,
  \textbf{7}, 2353--2357\relax
\mciteBstWouldAddEndPuncttrue
\mciteSetBstMidEndSepPunct{\mcitedefaultmidpunct}
{\mcitedefaultendpunct}{\mcitedefaultseppunct}\relax
\EndOfBibitem
\bibitem[Limpanuparb \emph{et~al.}(2012)Limpanuparb, Hollett, and Gill]{RO6}
T.~Limpanuparb, J.~W. Hollett and P.~M.~W. Gill, \emph{J. Chem. Phys.}, 2012,
  \textbf{136}, 104102\relax
\mciteBstWouldAddEndPuncttrue
\mciteSetBstMidEndSepPunct{\mcitedefaultmidpunct}
{\mcitedefaultendpunct}{\mcitedefaultseppunct}\relax
\EndOfBibitem
\bibitem[Limpanuparb \emph{et~al.}(2013)Limpanuparb, Milthorpe, Rendell, and
  Gill]{RO7}
T.~Limpanuparb, J.~Milthorpe, A.~Rendell and P.~Gill, \emph{J. Chem. Theory
  Comput.}, 2013, \textbf{9}, 863--867\relax
\mciteBstWouldAddEndPuncttrue
\mciteSetBstMidEndSepPunct{\mcitedefaultmidpunct}
{\mcitedefaultendpunct}{\mcitedefaultseppunct}\relax
\EndOfBibitem
\bibitem[Limpanuparb \emph{et~al.}(2014)Limpanuparb, Milthorpe, and
  Rendell]{RO8}
T.~Limpanuparb, J.~Milthorpe and A.~Rendell, \emph{J. Comput. Chem.}, 2014,
  \textbf{35}, In press\relax
\mciteBstWouldAddEndPuncttrue
\mciteSetBstMidEndSepPunct{\mcitedefaultmidpunct}
{\mcitedefaultendpunct}{\mcitedefaultseppunct}\relax
\EndOfBibitem
\bibitem[Limpanuparb(2012)]{ROThesis}
T.~Limpanuparb, \emph{Applications of Resolutions of the Coulomb Operator in
  Quantum Chemistry}, PhD dissertation, Australian National University,
  http://hdl.handle.net/1885/8879, 2012\relax
\mciteBstWouldAddEndPuncttrue
\mciteSetBstMidEndSepPunct{\mcitedefaultmidpunct}
{\mcitedefaultendpunct}{\mcitedefaultseppunct}\relax
\EndOfBibitem
\bibitem[Gautschi(1967)]{gautschi1967}
W.~Gautschi, \emph{SIAM Rev.}, 1967, \textbf{9}, 24--82\relax
\mciteBstWouldAddEndPuncttrue
\mciteSetBstMidEndSepPunct{\mcitedefaultmidpunct}
{\mcitedefaultendpunct}{\mcitedefaultseppunct}\relax
\EndOfBibitem
\bibitem[Olver and Sookne(1972)]{olver1972}
F.~W.~J. Olver and D.~J. Sookne, \emph{Math. Comput.}, 1972, \textbf{26},
  941--947\relax
\mciteBstWouldAddEndPuncttrue
\mciteSetBstMidEndSepPunct{\mcitedefaultmidpunct}
{\mcitedefaultendpunct}{\mcitedefaultseppunct}\relax
\EndOfBibitem
\bibitem[Olver(1964)]{olver1964}
F.~W.~J. Olver, \emph{Math. Comput.}, 1964, \textbf{18}, 65--74\relax
\mciteBstWouldAddEndPuncttrue
\mciteSetBstMidEndSepPunct{\mcitedefaultmidpunct}
{\mcitedefaultendpunct}{\mcitedefaultseppunct}\relax
\EndOfBibitem
\bibitem[{Lord Rayleigh (J. W. Strutt)}(1910)]{lordRayleigh1910}
{Lord Rayleigh (J. W. Strutt)}, \emph{Proc. R. Soc. A}, 1910, \textbf{84},
  25--46\relax
\mciteBstWouldAddEndPuncttrue
\mciteSetBstMidEndSepPunct{\mcitedefaultmidpunct}
{\mcitedefaultendpunct}{\mcitedefaultseppunct}\relax
\EndOfBibitem
\bibitem[Press \emph{et~al.}(1992)Press, Teukolsky, Vetterling, and
  Flannery]{NR2nd1992}
W.~H. Press, S.~A. Teukolsky, W.~T. Vetterling and B.~P. Flannery,
  \emph{Numerical {Recipes} in {C}. The art of scientific computing}, Cambridge
  University Press, 2nd edn., 1992, vol.~1\relax
\mciteBstWouldAddEndPuncttrue
\mciteSetBstMidEndSepPunct{\mcitedefaultmidpunct}
{\mcitedefaultendpunct}{\mcitedefaultseppunct}\relax
\EndOfBibitem
\bibitem[Press(2007)]{NR3rd2007}
W.~H. Press, \emph{Numerical {Recipes} 3rd edition: The art of scientific
  computing}, Cambridge University Press, 2007\relax
\mciteBstWouldAddEndPuncttrue
\mciteSetBstMidEndSepPunct{\mcitedefaultmidpunct}
{\mcitedefaultendpunct}{\mcitedefaultseppunct}\relax
\EndOfBibitem
\bibitem[Galler(1960)]{galler1960}
G.~Galler, \emph{Commun. ACM}, 1960, \textbf{3}, 353\relax
\mciteBstWouldAddEndPuncttrue
\mciteSetBstMidEndSepPunct{\mcitedefaultmidpunct}
{\mcitedefaultendpunct}{\mcitedefaultseppunct}\relax
\EndOfBibitem
\bibitem[Herndon(1961)]{herndon1961}
J.~R. Herndon, \emph{Commun. ACM}, 1961, \textbf{4}, 178--179\relax
\mciteBstWouldAddEndPuncttrue
\mciteSetBstMidEndSepPunct{\mcitedefaultmidpunct}
{\mcitedefaultendpunct}{\mcitedefaultseppunct}\relax
\EndOfBibitem
\bibitem[Wiggins and Saito(1971)]{wiggins1971}
R.~A. Wiggins and M.~Saito, \emph{Bull. Seismol. Soc. Am.}, 1971, \textbf{61},
  375--381\relax
\mciteBstWouldAddEndPuncttrue
\mciteSetBstMidEndSepPunct{\mcitedefaultmidpunct}
{\mcitedefaultendpunct}{\mcitedefaultseppunct}\relax
\EndOfBibitem
\bibitem[Braithwaite(1973)]{braithwaite1973}
W.~Braithwaite, \emph{Comput. Phys. Commun.}, 1973, \textbf{5}, 390--394\relax
\mciteBstWouldAddEndPuncttrue
\mciteSetBstMidEndSepPunct{\mcitedefaultmidpunct}
{\mcitedefaultendpunct}{\mcitedefaultseppunct}\relax
\EndOfBibitem
\bibitem[Schneider \emph{et~al.}(2010)Schneider, Segura, Gil, Guan, and
  Bartschat]{schneider2010}
B.~I. Schneider, J.~Segura, A.~Gil, X.~Guan and K.~Bartschat, \emph{Comput.
  Phys. Commun.}, 2010, \textbf{181}, 2091--2097\relax
\mciteBstWouldAddEndPuncttrue
\mciteSetBstMidEndSepPunct{\mcitedefaultmidpunct}
{\mcitedefaultendpunct}{\mcitedefaultseppunct}\relax
\EndOfBibitem
\bibitem[Smith \emph{et~al.}(1981)Smith, Olver, and Lozier]{smith1981}
J.~Smith, F.~Olver and D.~W. Lozier, \emph{ACM T. Math. Software}, 1981,
  \textbf{7}, 93--105\relax
\mciteBstWouldAddEndPuncttrue
\mciteSetBstMidEndSepPunct{\mcitedefaultmidpunct}
{\mcitedefaultendpunct}{\mcitedefaultseppunct}\relax
\EndOfBibitem
\bibitem[Lozier and Smith(1981)]{lozier1981}
D.~W. Lozier and J.~Smith, \emph{ACM T. Math. Software}, 1981, \textbf{7},
  141--146\relax
\mciteBstWouldAddEndPuncttrue
\mciteSetBstMidEndSepPunct{\mcitedefaultmidpunct}
{\mcitedefaultendpunct}{\mcitedefaultseppunct}\relax
\EndOfBibitem
\bibitem[Olver and Smith(1983)]{olver1983}
F.~Olver and J.~Smith, \emph{J. Comput. Phys.}, 1983, \textbf{51},
  502--518\relax
\mciteBstWouldAddEndPuncttrue
\mciteSetBstMidEndSepPunct{\mcitedefaultmidpunct}
{\mcitedefaultendpunct}{\mcitedefaultseppunct}\relax
\EndOfBibitem
\bibitem[Libbrecht(1985)]{libbrecht1985}
K.~G. Libbrecht, \emph{Sol. Phys.}, 1985, \textbf{99}, 371--373\relax
\mciteBstWouldAddEndPuncttrue
\mciteSetBstMidEndSepPunct{\mcitedefaultmidpunct}
{\mcitedefaultendpunct}{\mcitedefaultseppunct}\relax
\EndOfBibitem
\bibitem[Holmes and Featherstone(2002)]{holmes2002}
S.~A. Holmes and W.~E. Featherstone, \emph{J. Geodesy}, 2002, \textbf{76},
  279--299\relax
\mciteBstWouldAddEndPuncttrue
\mciteSetBstMidEndSepPunct{\mcitedefaultmidpunct}
{\mcitedefaultendpunct}{\mcitedefaultseppunct}\relax
\EndOfBibitem
\bibitem[Jekeli \emph{et~al.}(2007)Jekeli, Lee, and Kwon]{jekeli2007}
C.~Jekeli, J.~K. Lee and J.~H. Kwon, \emph{J. Geodesy}, 2007, \textbf{81},
  603--615\relax
\mciteBstWouldAddEndPuncttrue
\mciteSetBstMidEndSepPunct{\mcitedefaultmidpunct}
{\mcitedefaultendpunct}{\mcitedefaultseppunct}\relax
\EndOfBibitem
\bibitem[Galassi \emph{et~al.}(2009)Galassi, Davies, Theiler, Gough, Jungman,
  Alken, Booth, and Rossi]{GNU2009}
M.~Galassi, J.~Davies, J.~Theiler, B.~Gough, G.~Jungman, P.~Alken, M.~Booth and
  F.~Rossi, \emph{GNU scientific library reference manual}, Network Theory
  Ltd., 2009\relax
\mciteBstWouldAddEndPuncttrue
\mciteSetBstMidEndSepPunct{\mcitedefaultmidpunct}
{\mcitedefaultendpunct}{\mcitedefaultseppunct}\relax
\EndOfBibitem
\bibitem[Schling(2011)]{BOOST2011}
B.~Schling, \emph{The Boost C++ libraries}, XML Press, 2011\relax
\mciteBstWouldAddEndPuncttrue
\mciteSetBstMidEndSepPunct{\mcitedefaultmidpunct}
{\mcitedefaultendpunct}{\mcitedefaultseppunct}\relax
\EndOfBibitem
\bibitem[Sloan(2013)]{sloan2013}
P.-P. Sloan, \emph{J. Comput. Graph. Techniques}, 2013, \textbf{2},
  84--90\relax
\mciteBstWouldAddEndPuncttrue
\mciteSetBstMidEndSepPunct{\mcitedefaultmidpunct}
{\mcitedefaultendpunct}{\mcitedefaultseppunct}\relax
\EndOfBibitem
\end{mcitethebibliography}
\bibliographystyle{rsc}

\end{document}